\documentclass[twocolumn,aps,prl,showpacs,amsmath,amssymb,nofootinbib]{revtex4} \usepackage{hyperref}
\usepackage{graphicx}

\newcommand{\Ueff}{U_{\textrm{eff}}}
\newcommand{\Fa}{F^{(0)}}
\newcommand{\Fb}{F^{(2)}}
\newcommand{\Fc}{F^{(4)}}

\newcommand{\og}{1/\gamma}
\newcommand{\na}{\bar{n}}
\newcommand{\EN}{E_n}

\begin{document}

\title{Valence-skipping and negative-U in the d-band\\ 
from repulsive local Coulomb interaction}

\author{Hugo U.~R.~Strand}
\email{hugo.strand@physics.gu.se}
\affiliation{University of Gothenburg, SE-412 96 Gothenburg, Sweden} 
\date{\today} 

\begin{abstract}

 We show that repulsive local Coulomb interaction alone can drive valence-skipping charge disproportionation in the degenerate d-band, resulting in effective negative-U. This effect is shown to originate from anisotropic orbital-multipole scattering, and occurs only for $d^1$, $d^4$, $d^6$ and $d^9$ fillings (and their immediate surroundings).
 Explicit boundaries for valence-skipping are derived and the paramagnetic phase diagram for $d^4$ and $d^6$ is calculated. We also establish that the valence-skipping metal is very different, in terms of its local valence distribution, compared to the atomic-like Hund's metal.
 These findings explains why transition metal compounds with the aforementioned d-band fillings are more prone to valence-skipping charge order and anomalous superconductivity.

\end{abstract}
\maketitle
\makeatletter
\let\toc@pre\relax
\let\toc@post\relax
\makeatother

 While going up in an elevator, have you ever caught yourself staring at the numbers flying by? The eleventh, the twelfth, and then all of a sudden the fourteenth floor. Valence-skipping elements are just like elevators, except they avoid certain valence states rather than the thirteenth floor.
 The most prominent valence-skippers are the post-transition metals, Tl, Bi, Sb, etc., who display missing valences in many of their compounds \cite{Varma:1988pi}. In general valence-skipping is driven by a negative effective Coulomb repulsion $\Ueff$, but the mechanism causing this is debated.
 Anderson \cite{Anderson:1975bf} showed that static lattice relaxation can drive $\Ueff < 0$. However Varma \cite{Varma:1988pi} noted that even free atoms have reduced $\Ueff$ in the unfavorable valences, and proposed an intra-atomic electronic mechanism.
 But more recently the electronic route has been discredited for these elements, in favor of the lattice relaxation mechanism \cite{Harrison:2006if}.

 Apart from the post-transition metals, valence-skipping has also been observed in transition metals \cite{Varma:1988pi} with a $d^n \rightarrow d^{n-1}$+$d^{n+1}$ type of charge disproportionation.
 Experimentally the most evident examples are the iron compounds (La,Ca)FeO$_3$ \cite{Liang:2005xz, Matsuno:2002jb}, (La,Sr)FeO$_3$ \cite{Matsuno:1999wm, Matsuno:2002jb}, and Sr$_3$Fe$_2$O$_7$ \cite{JPSJ.69.2767}, where M\"ossbauer spectroscopy has established valence-skipping Fe$^{4+} \rightarrow$ Fe$^{3+}$+Fe$^{5+}$, ($d^4 \rightarrow d^3 $+$ d^5$) charge-order, even in the absence of lattice relaxation \cite{Liang:2005xz, JPSJ.69.2767}.
 Theoretically Katayama-Yoshida and Zunger \cite{Katayama-Yoshida:1985lo} showed that effective monopole screening of intra-atomic interactions indeed can give rise to $\Ueff < 0$ in transition metal impurities. This idea has been used to explain the charge-order in YNiO$_3$ \cite{Mazin:2007fd} within a two-band $e_g$ model. But the complete d-band still deserves more attention.

 The fact that some authors even refer to valence-skipping as ``mysterious'' \cite{Hase:2007wq, Moskvin:2009hs}, show the need for better understanding of the underlying mechanism behind this phenomenon and its systematics.
 The resulting negative-U model however, has been studied extensively, and shown to drive both charge-order and anomalous superconductivity \cite{Micnas:1990wb}. So unveiling the mysteries of valence-skipping could pave the way for more exotic physics.

 In this Letter we show that valence-skipping in the degenerate d-band is driven by the higher orbital-multipole part of the repulsive intra-atomic Coulomb interaction. This effect is found to be limited to the particular fillings $d^1$, $d^4$, $d^6$, and $d^9$, and their immediate surroundings. From a multiplet analysis we derive explicit bounds for valence-skipping, and finally the emerging anomalous valence fluctuations in the paramagnetic metal are studied.


 Let us begin by constructing a minimal model for the correlated d-band. We assume that the Coulomb interaction is local and rotationally invariant, a good first approximation for transition metals \cite{Vaugier:2012kx}. Under this assumption, the interaction is exactly given by the Slater-Condon angular-momentum expansion, and the Slater-integrals $F^{(0)}$, $F^{(2)}$ and $F^{(4)}$ \cite{Rudzikas:2007zr}. For the electron hopping we use a degenerate semi-circular density of states, and take the half-bandwidth as our unit of energy.

 In general the local interaction describes electron-pair scattering between local two-particle states, and rotational invariance ensure that these processes conserve local total orbital momentum $L$ and total spin $S$. As we are going to see, anisotropic orbital-multipole scattering (i.e.~for $L > 0$) has an intrinsic connection to valence-skipping. To make this clear we now seek to isolate this contribution to the interaction.

 Within the Slater-Condon interaction, the $F^{(0)}$ term is a density-density interaction with isotropic scattering, while the $F^{(2)}$ and $F^{(4)}$ terms have different scattering strengths for all $L$ and $S$. Interestingly their orbital-multipole anisotropies cancel out when $F^{(4)}/F^{(2)} = 9/5$. This corresponds to a Laporte-Platt degenerate point of the Slater-Condon interaction with large accidental degeneracies of multiplets \cite{Judd:1984vn}.

 Spurred by this observation we investigated the Slater-Condon interaction in detail \cite{Strand:2013ve}, and found that in this point it simplifies to the rotationally invariant Kanamori interaction \cite{Mizokawa:1995kx}, having the compact form
\begin{equation}
  \hat{H} = (U-3J) \hat{N}(\hat{N}-1)/2 + J(\hat{Q}^2-\hat{S}^2) 
  \, , \label{eq:SQKanamori}
\end{equation}
 where $\hat{N}$ is the total number operator, and $\hat{S}^2$ and $\hat{Q}^2$ are the total spin and quasi-spin operators \cite{Rudzikas:2007zr}.

 The Kanamori interaction and the roles of its coupling parameters, the Hubbard $U$, and Hund's rule $J$ have already been studied extensively \cite{2012arXiv1207.3033G}. This makes the reduction of the Slater-Condon interaction (at the Laporte-Platt degenerate point $F^{(4)}/F^{(2)} = 9/5$) very interesting. At this point $U$ and $J$ alone can be used to determine $F^{(k)}$.

 But let us first establish our claim that anisotropic orbital-multipole scattering is indeed missing in Eq.~(\ref{eq:SQKanamori}). We need not to worry about the density-density interaction giving isotropic scattering (like the $\Fa$ term).
 So all non-trivial scattering in Eq.~(\ref{eq:SQKanamori}) stems from $J(\hat{Q}^2 - \hat{S}^2)$, where $\hat{S}^2$ (acting in spin space) do not differentiate between orbital angular-momentum channels $L$ directly. The quasi-spin operator $\hat{Q}^2$ however, do differentiate in $L$, but scatters only in the monopole channel ($L=0$) \cite{Rudzikas:2007zr}. This proves our point; the Slater-Condon interaction, at the Laporte-Platt degenerate point $\Fc/\Fb = 9/5$, is free from anisotropic orbital-multipole interactions. Here on we will refer to these interactions as simply ``multipole-interactions''.

 Guided by our findings we propose the following re-parametrization of the Slater-Condon interaction
\begin{equation}
  \Fa = U - \frac{8}{5} J, \,
  \Fb = 49 \left( \frac{1}{\gamma} + \frac{1}{7} \right) J, \,
  \Fc = \frac{63}{5} J,
  \label{eq:SCparam}
\end{equation}
using $U$, $J$ and $1/\gamma$ that control the relative strength of the multipole interactions.
 A cautionary remark is in place, the multipole parametrization is arbitrary
 (Ref.~\cite{Marel:1988uq} uses another equivalent form), and the choice of $\og$ in Eq.~(\ref{eq:SCparam}) is a matter of taste.\footnote{ Our choice is motivated by the simple form Eq.~(\ref{eq:SCparam}) takes in terms of the Racah parameters, $A = U - 3J$, $B = J/\gamma$, $C = J$.
}
 However the Kanamori limit, without multipole terms, is well defined by $1/\gamma = 0$. In what follows we set $1/\gamma = 1/4$, which corresponds to $F^{(4)}/F^{(2)} \approx 0.65$, in the relevant regime for the transition metals \cite{Groot:1990vn}. 


 We are now in a position to start our study of the d-band model. Much can in fact be learned in the limit of strong interactions, where the system turns in to an ensemble of isolated atoms with known $n$-electron ground state energies $\EN$.
 For the ensemble with integer average filling $\na$ the obvious ground state candidate is the homogeneous state with energy $E_{\na}$. But there is also the possibility of phase-separated mixtures of atomic states with $n_1$ and $n_2$ electrons.
 In general such a mixture has the energy $E_{n_1, n_2}^{(\na)} = E_{n_1} + (E_{n_2}-E_{n_1})(\na-n_1)/(n_2-n_1)$ assuming $n_1 < \na < n_2$.

\begin{figure}
  \includegraphics[scale=1]{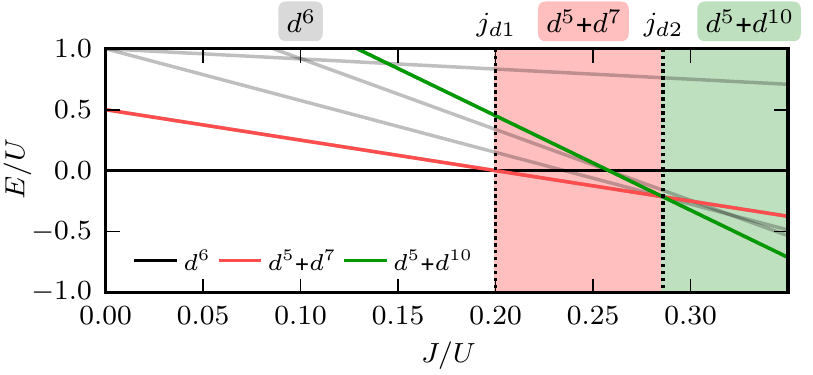}
  \caption{\label{fig:EnergyCrossings} (color online).
  Ensemble energies as a function of $J/U$ for all mixed-valence states (gray lines) relative to the atomic $d^6$ ground state (black line) for $1/\gamma = 1/4$. The energy crossings $j_{d1}$ and $j_{d2}$  (dotted lines) in to the $d^5$+$d^7$ valence-skipping (red line) and $d^5$+$d^{10}$ split-valence (green line) phases are indicated.}
\end{figure}

\begin{figure}
  \includegraphics[scale=1]{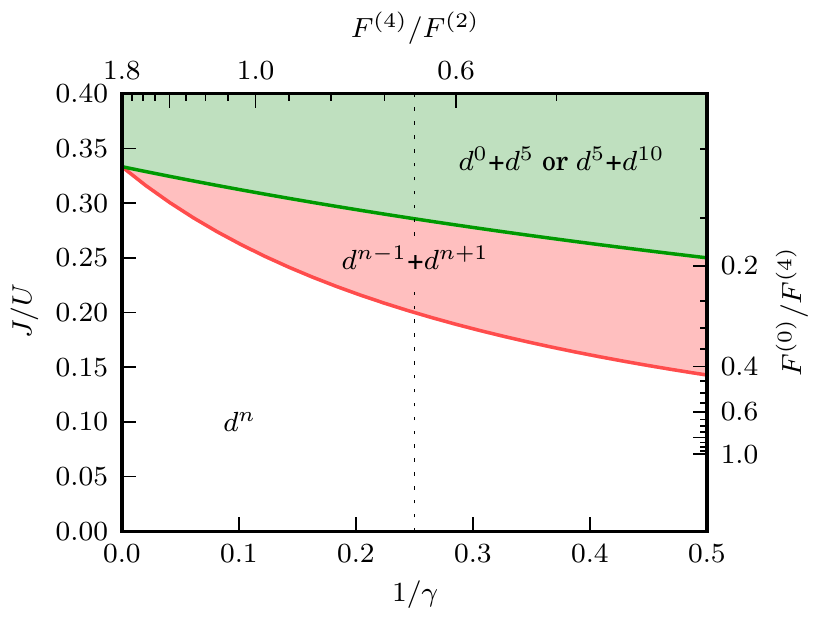}
  \caption{\label{fig:EnsemblePhases} (color online).
  Ensemble phase diagram in the ($J/U$, $1/\gamma$) and ($\Fc/\Fb$, $\Fa/\Fc$) plane for integer average fillings $n$. The valence-skipping $d^{n-1}$+$d^{n+1}$ phase is only present for $n = 1,4,6$ and $9$. The dotted line corresponds to $1/\gamma = 1/4$.}
\end{figure}

 We have compared all candidate states for every integer $\na$ and located the ground state crossings as a function of $J/U$ and $\og$, see Fig.~\ref{fig:EnergyCrossings} for an example. 
 We find that (like in Fig.~\ref{fig:EnergyCrossings}) the valence-skipping $d^{\na-1}$+$d^{\na+1}$ state is the ensemble ground state in the range, $j_{d1} < J/U < j_{d2}$, with $\og$ dependent bounds $j_{d1} = 1/(3 + 8/\gamma)$ and $j_{d2} = 1/(3 + 2/\gamma)$, but only for $\na = 1, 4, 6$ and $9$. For the other integer $\na$ the $d^{\na-1}$+$d^{\na+1}$ state never becomes the ground state. When $J/U > j_{d2}$, the ensemble has a split-valence type of ground state for all $\na$, composed by $d^{0}$+$d^{5}$ for $\na<5$ and $d^{5}$+$d^{10}$ for $\na > 5$ (as in Fig.~\ref{fig:EnergyCrossings}). The final phase diagram is shown in Fig.~\ref{fig:EnsemblePhases}.

 With this background we can understand the connection between valence-skipping and effective negative-U. The effective Hubbard repulsion $\Ueff$ is given by \cite{Katayama-Yoshida:1985lo}
\begin{equation}
  \Ueff = E_{\na+1}+E_{\na-1}-2E_{\na} = 2(E^{(\na)}_{\na-1, \na+1}-E_{\na}) \, ,
  \label{eq:Ueff}
\end{equation}
 and $\Ueff < 0$ occurs only for concave series $E_{\na-1}$, $E_{\na}$, $E_{\na+1}$. In the case of a valence-skipping ensemble ground state $d^{\na-1}$+$d^{\na+1}$ we are guaranteed that $E^{(\na)}_{\na-1, \na+1} < E_{\na}$, and Eq.~(\ref{eq:Ueff}) directly gives $\Ueff < 0$. 

 But what is now the role of the multipole interactions? From the ensemble-phase diagram (Fig.~\ref{fig:EnsemblePhases}) it is clear that the multipole interaction strength $\og$ directly controls the extent of the valence-skipping phase, and in the limit $\og \rightarrow 0$ this phase disappears. We conclude that the valence-skipping ground state is realized by the multipole interactions. 

\begin{figure}
  \includegraphics[scale=1]{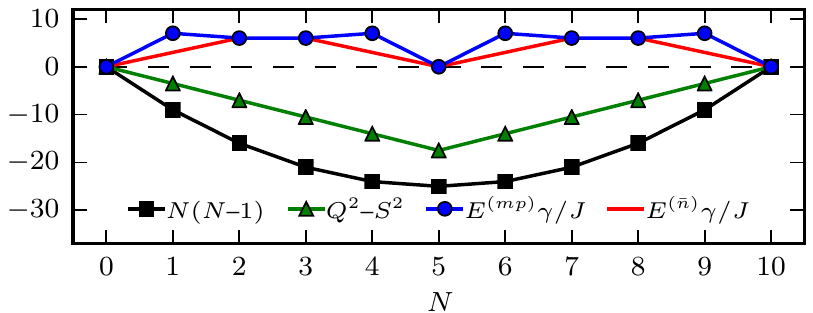}
  \caption{\label{fig:EnergyContribs} (color online).
  Atomic ground state energy contributions for all $N$-electron fillings; density-density (squares), spin and quasi-spin scattering (triangles), and the multipole energies $E^{(\textrm{mp})}$ (circles), up to (irrelevant) linear shifts $\mu N$. Note that only $E^{(\textrm{mp})}$ is locally concave, and only for $N=1,4,6,9$, where the valence-skipping state has a lower multipole contribution $E^{(\na)}$ (red lines).}
\end{figure}

 To understand why the effect is limited to only certain fillings we decompose the atomic ground state energies $E_n$ in Kanamori and multipole contributions. As seen in Fig.~\ref{fig:EnergyContribs} the isotropic and mono-pole terms are convex (as long as $U - 3J > 0$). However the multipole energy $E^{(\textrm{mp})}_n = E_n - (U-3J)N(N-1)/2 - J(\langle \hat{Q}^2 \rangle - \langle \hat{S}^2 \rangle)$ is locally concave, but only for the special fillings $n = 1,4,6,9$ and can therefore give $\Ueff < 0$ for sufficiently large $J/\gamma$. Because of this we will henceforth denote these fillings as ``multipole-active''.
 From Fig.~\ref{fig:EnergyContribs} it is also clear that the valence-skipping $d^{n-1}$+$d^{n+1}$ ground state at multipole-active filling $n$, is stable with respect to doping in the whole range $n-1 < \na < n+1$ of $d^{\na}$ fillings.

 Let us close the discussion of the ensemble by recasting the valence-skipping criteria $J/U > j_{d1}$ in terms of $F^{(k)}$
\begin{equation}
  \frac{\Fc}{\Fb} 
  \frac{ \frac{\Fa}{\Fc}-\frac{1}{9} }{ \frac{9}{5}-\frac{\Fc}{\Fb}  }
  < \frac{40}{441} \, .
  \label{eq:FkReq}
\end{equation}
 As $\Fc/\Fb$ varies weakly within the transition metals, fulfillment of Eq.~(\ref{eq:FkReq}) is mainly driven by ligand induced effective monopole screening of $F^{(0)}$ \cite{Katayama-Yoshida:1985lo}.


 With these insights we leave the strong coupling limit and consider the full model with its competition between itineracy and local interactions. The ground state is calculated using the variational Gutzwiller method \cite{Fabrizio:2007aa, Lanata:2008ly, Lanata:2012lr}, previously shown to give phase diagrams in qualitative agreement with dynamical mean field theory \cite{Huang:2012hc}. We limit the discussion to translationally invariant paramagnetic wave-functions, employing the most general variational ansatz with the symmetry of our model.

\begin{figure}
  \includegraphics[scale=1]{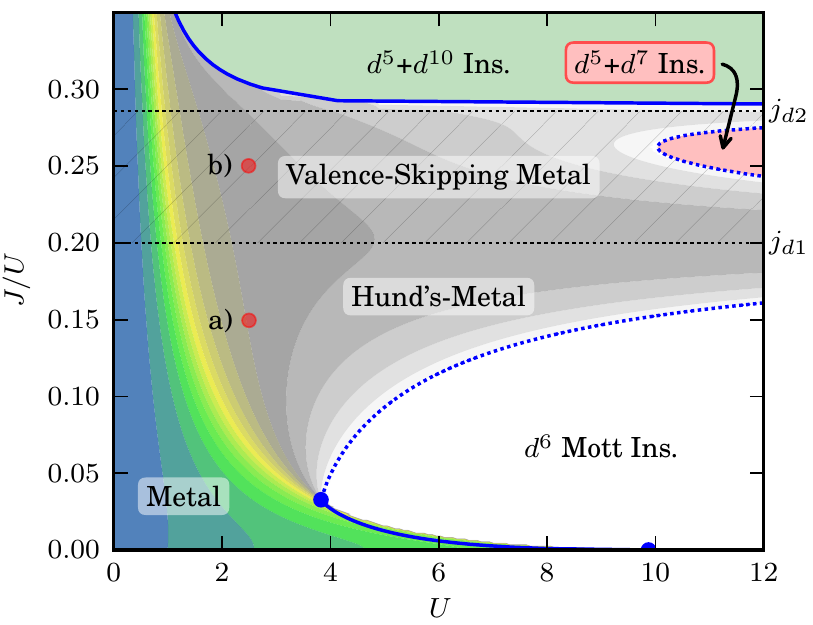}
  \caption{\label{fig:PhaseDiag} (color online).
  Phase diagram for $d^6$, with $1/\gamma = 1/4$, showing the contours of the local entanglement entropy of the metallic state, and the metal-insulator phase-boundaries (blue lines), first and second order transitions are indicated (solid and dotted lines respectively).}
\end{figure}

 Here we report results for $d^6$ (particle-hole symmetric to $d^4$), whose phase-boundaries are shown in Fig.~\ref{fig:PhaseDiag}, together with the local entanglement entropy contours \cite{Amico:2008vn} of the metal. The low $J/U$ region ($J/U < j_{d1}$) agrees qualitatively with the three-band Kanamori model \cite{Medici:2011kl, Medici:2011qf} and will not be discussed further. A quantitative comparison is left for future works \cite{Strand:2013ve}.

 Our current interest lies in the Hund's-metal \cite{2012arXiv1207.3033G} and valence-skipping regimes ($j_{d1} \lesssim J/U \lesssim j_{d2}$). In general for fixed $J/U$ there is a critical coupling $U = U_c$ where the metal-insulator transition occurs.
 But as seen in Fig.~\ref{fig:PhaseDiag}, the $U_c$ of the Hund's-metal grows with increased $J/U$, and when $J/U \rightarrow j_{d1}$ it diverges ($U_c \rightarrow \infty$). At this point, $J/U = j_{d1}$, the metallicity prevails for any $U$ because the energy-cost for charge fluctuations is zero, $\Ueff(j_{d1}) = 0$. When entering the valence-skipping regime ($j_{d1} < J/U < j_{d2}$), $U_c$ becomes finite again as a reentrant valence-skipping $d^5$+$d^7$ insulator emerges. Yet, approaching the upper boundary $J/U \rightarrow j_{d2}$, $U_c$ diverges again. Further increasing $J/U$ rapidly reduces $U_c$ in favor of a split-valence $d^{5}$+$d^{10}$ insulator.


 How is then the metal influenced by the change in the ensemble ground state from $d^6$ to $d^5$+$d^7$? In terms of local valences the single-particle hopping in the metal generates a distribution of adjacent valences. This distribution however, is strongly dependent on the intra-atomic interaction. 

\begin{figure}
  \includegraphics[scale=1]{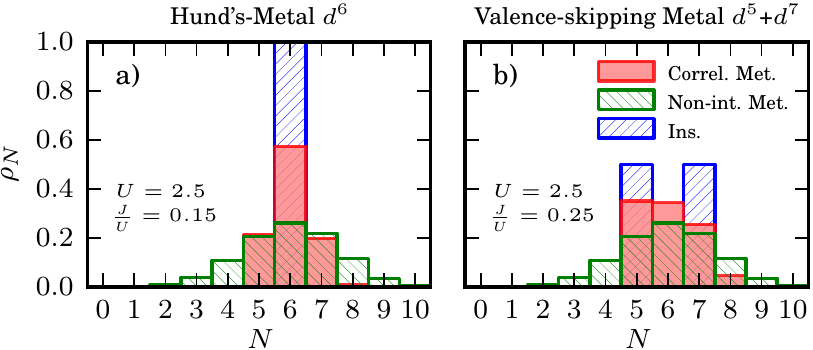}
  \caption{\label{fig:Histograms} (color online).
 Histograms of valence weights $\rho_N$ as a function of filling $N$ for the points marked in Fig.~\ref{fig:PhaseDiag}, with $1/\gamma = 1/4$. The correlated metal (red) is shown together with the corresponding U=0 non-interacting metal (green), and the $U \rightarrow \infty$ insulator (blue), for a) the Hund's-metal, and b) the valence-skipping $d^5$+$d^7$ metal.}
\end{figure}

 To investigate this we compute the reduced local many-body density matrix $\hat{\rho}$ \cite{Amico:2008vn}, and calculate the valence weights $\rho_N$ as traces of $\hat{\rho}$ in every $N$-electron subspace. The valence-distributions $\rho_N$ for the Hund's and valence-skipping metals are shown in Fig.~\ref{fig:Histograms}, at the points marked out in the phase diagram (Fig.~\ref{fig:PhaseDiag}).
 In each case $\rho_N$ for the corresponding non-interacting metal ($U=0$) and insulator ($U\rightarrow \infty$) are shown for comparison.

 The Hund's-metal in Fig.~\ref{fig:Histograms}a, has an atomic-like valence distribution that is substantially narrower compared to the non-interacting metal. Most of the weight is concentrated in the range $N = 5 - 7$, with a strong prevalence towards the total average valence $\bar{N} = 6$.
 Turning to Fig.~\ref{fig:Histograms}b and the valence-skipping metal, we find the same narrowing down of the distribution, but without any certain valence prevalence. Thus, in comparison to the Hund's-metal, there is a substantial reduction of the average $d^6$ valence. This type of reduction is the hallmark of the anomalous valence fluctuations driven by the effective negative-U in the valence-skipping region.

 We have shown that the local multipole interactions drastically reduce the effective Hubbard repulsion $\Ueff$ in the d-band, even making it possible to reach negative-U ($\Ueff < 0$). Moreover this multipole reduction is only obtained for four out of ten possible integer d-band fillings, namely $d^1$, $d^4$, $d^6$ and $d^9$.
 Admittedly we have used an over-simplistic model of the d-band. But the valence-skipping active fillings is a fundamental property of the Coulomb interaction, and apply to the entire class of transition metals.

 Experimentally valence-skipping is most clearly observed when accompanied by charge-order and $\Ueff < 0$, as in the iron $d^4$ compounds discussed in the introduction \cite{Liang:2005xz, Matsuno:2002jb, Matsuno:1999wm, JPSJ.69.2767}, and noble-metal $d^9$ systems such as CsAuI$_3$ \cite{Hafner:1994zt}. 
 However multipole-reduced but positive $\Ueff \gtrsim 0$ also generate valence-skipping in terms of polarons at elevated temperatures $T \gtrsim \Ueff$. This type of thermally induced valence-skipping has been used to explain the polaronic conduction in $d^6$ (La,Ca)CoO$_3$ \cite{Sehlin:1995gd} and $d^4$ (La,Ca)MnO$_3$ \cite{Hundley:1997ta}. 
 For the $d^1$ filling \cite{Varma:1988pi} some of the candidate transition metal complex-oxide compounds are not even thermodynamically stable, e.g., La$_2$V$_2$O$_7$ phase-separate directly to LaVO$_3$ and LaVO$_4$ ($d^1 \rightarrow d^0$+$d^2$) \cite{Yokokawa:1992tn}.

 So returning to the propositions of Anderson \cite{Anderson:1975bf} and Varma \cite{Varma:1988pi}, we conclude that for multipole-active fillings the electron interaction can drive valence-skipping even in absence of lattice relaxation. One such example is La$_{1/2}$Ca$_{1/2}$FeO$_3$ \cite{Liang:2005xz} that charge-orders to $3(d^{3.5}) \rightarrow 2(d^{3})$+$d^{5}$. While in other cases both multipole-interactions and static lattice relaxation combine to give $\Ueff < 0$.
 Note that, even thought the rules for multipole-active fillings were derived for the degenerate d-band model, they remain applicable also in weak crystal-fields. However in the limit of strong crystal-fields they break down, like in YNiO$_3$ that show $d^7$ valence-skipping charge-order isolated to the crystal-field split $e_g$-states, $t_{2g}^6 e_g^1 \rightarrow t_{2g}^6 e_g^0 + t_{2g}^6 e_g^2$ \cite{Mazin:2007fd}.

 From the Gutzwiller calculations it is clear that multipole interactions also affect the metallic state. 
 The importance of the metallic valence distribution has been discussed in a recent study of SrCoO$_3$ using the Kanamori interaction  \cite{PhysRevLett.109.117206}. Here a followup study also including multipole-interactions through the Slater-Condon interaction would be very interesting.

 Although valence-skipping is experimentally most evident in charge ordered compounds, negative-U is also a potential electron-pairing mechanism for superconductivity \cite{Micnas:1990wb}. Therefore it is worth noting that the cuprate ($d^9$:\,Cu$^{2+}$), ruthanate ($d^4$:\,Ru$^{4+}$), and iron pnictide and chalcogenide ($d^6$:\,Fe$^{2+}$) superconductors, all have multipole-active d-band fillings.


 To conclude we have shown that, in the vicinity of the multipole-active fillings $d^1$, $d^4$, $d^6$, and $d^9$, the multipole part of the Slater-Condon interaction can alone drive valence-skipping $d^n \rightarrow d^{n-1}$+$d^{n+1}$ and negative-U in the degenerate d-band. Further more the valence fluctuations in the valence-skipping metal are drastically different compared to the atomic-like Hund's metal. None of these effects are captured by the Kanamori interaction, due to its lack of anisotropic multipole interactions.

\begin{acknowledgments}

 I would like to acknowledge Bo Hellsing and Mats Granath for fruitful discussions and careful reading of the manuscript and Paul Erhart for discussions on valence states. This work was supported by the Mathematics - Physics Platform ($\mathcal{MP}^{\textsf{2}}$) at the University of Gothenburg, and the Royal and Hvitfeldtska foundation through a visiting stipend for Jonsereds Mansion and Villa Martinsson where this manuscript was completed. The simulations were performed on resources provided by the Swedish National Infrastructure for Computing (SNIC) at Chalmers Centre for Computational Science and Engineering (C3SE) (project no.~01-11-297). 

\end{acknowledgments}

\bibliography{/Users/hugstr/Documents/Papers/DMFT_Biblography}

\end{document}